\documentclass{appolb}
\usepackage{graphicx,cite,hyperref}
\usepackage{epsfig}
\usepackage{amsmath, amssymb, slashed}
\usepackage{xcolor}
\usepackage{appendix}
\usepackage{dsfont}
\usepackage[normalem]{ulem}
\usepackage{bm}
\usepackage{amsfonts}
\newcommand{\ud}{\mathrm{d}}

\newcommand{\be}{\begin{equation}}
\newcommand{\ee}{\end{equation}}

\begin{document}
\title{\bf Pressure inside hadrons: criticism, conjectures, and all that}
\author{C\'edric Lorc\'e$^1$ and Peter Schweitzer$^2$
\address{\scriptsize 
    $^1$ CPHT, CNRS, \'Ecole polytechnique, Institut Polytechnique de Paris, 91120 Palaiseau, France\\ 
    $^2$ Department of Physics, University of Connecticut, Storrs, CT 06269, U.S.A.}}
\date{December 2024}
\maketitle
\begin{abstract}
The interpretation of the energy-momentum tensor form factor
$D(t)$ of hadrons in terms of pressure and shear force distributions
is discussed, concerns raised in the literature are reviewed, and ways 
to reconcile the concerns with the interpretation are indicated.
\end{abstract}
  
\section{Introduction}

This contribution is dedicated to the memory of Dmitry Diakonov, 
Victor Petrov and Maxim Polyakov whose lives and works strongly 
influenced the authors. Especially Maxim was a dear colleague,
mentor and friend for us. The purpose of this contribution
is to review the pressure interpretation of the $D$-term form 
factor, one of the most influential works of Maxim.

Continuous systems have well-defined mechanical properties
described by the theory of elasticity~\cite{LLv7}. In 2002 
Maxim Polyakov developed an appealing interpretation of hadronic
energy-momentum tensor (EMT) form factors under the assumption that the nucleon and nuclei
can be considered as ``continuous media''~\cite{Polyakov:2002yz}.
Especially the interpretation of the $D(t)$ form factor~\cite{Polyakov:1999gs} in terms of shear forces and pressure 
inside hadrons has attracted considerable interest~\cite{Schweitzer:2002nm,Goeke:2007fp,Goeke:2007fq,Cebulla:2007ei,Kim:2012ts,Mai:2012yc,Mai:2012cx,Jung:2013bya,Jung:2014jja,Pasquini:2014vua,Perevalova:2016dln,Hudson:2017xug,Hudson:2017oul,Burkert:2018bqq,Polyakov:2018zvc,Lorce:2018egm,Shanahan:2018nnv,Polyakov:2019lbq,Varma:2020crx,Neubelt:2019sou,Metz:2021lqv,Gegelia:2021wnj,Owa:2021hnj,Pefkou:2021fni,Lorce:2021xku,Freese:2022jlu,Freese:2022ibw,Freese:2022yur,Fu:2022rkn,Lorce:2022cle,Panteleeva:2023aiz,Amor-Quiroz:2023rke,Fu:2023ijy,Wang:2023bjp,Maynard:2024wyi,Wang:2024abv}, 
for a recent review see~\cite{Burkert:2023wzr}.
The interpretation has also raised concerns~
\cite{Jaffe:2020ebz,Freese:2021czn,Ji:2021mtz,Freese:2024rkr,Ji:2021mfb,Ji:2022exr,Fujita:2022jus,Czarnecki:2023yqd},
which are valuable as they help improve our physics understanding.

The intention of this work is to review the concerns raised in the literature,
collect thoughts and initiate a discussion. This work is organized as follows:
In Sec.~\ref{Sec-2:memories} we collect personal memories related to Maxim. 
In Sec.~\ref{Sec-3:interpret} we briefly review the interpretation proposed by Maxim~\cite{Polyakov:2002yz}. 
In Sec.~\ref{Sec-4:criticism} we review the criticism brought up in the literature, and 
collect thoughts in Secs.~\ref{Sec-5:comments-Breit-frame}-\ref{Sec-11:D-in-atoms} to argue that the proposed interpretation is physically sound 
and how the raised issues can be resolved, before concluding in 
Sec.~\ref{Sec-5:conclusions}.

\newpage

\begin{figure}[htb]
\centerline{
(a)
\includegraphics[height=5cm]{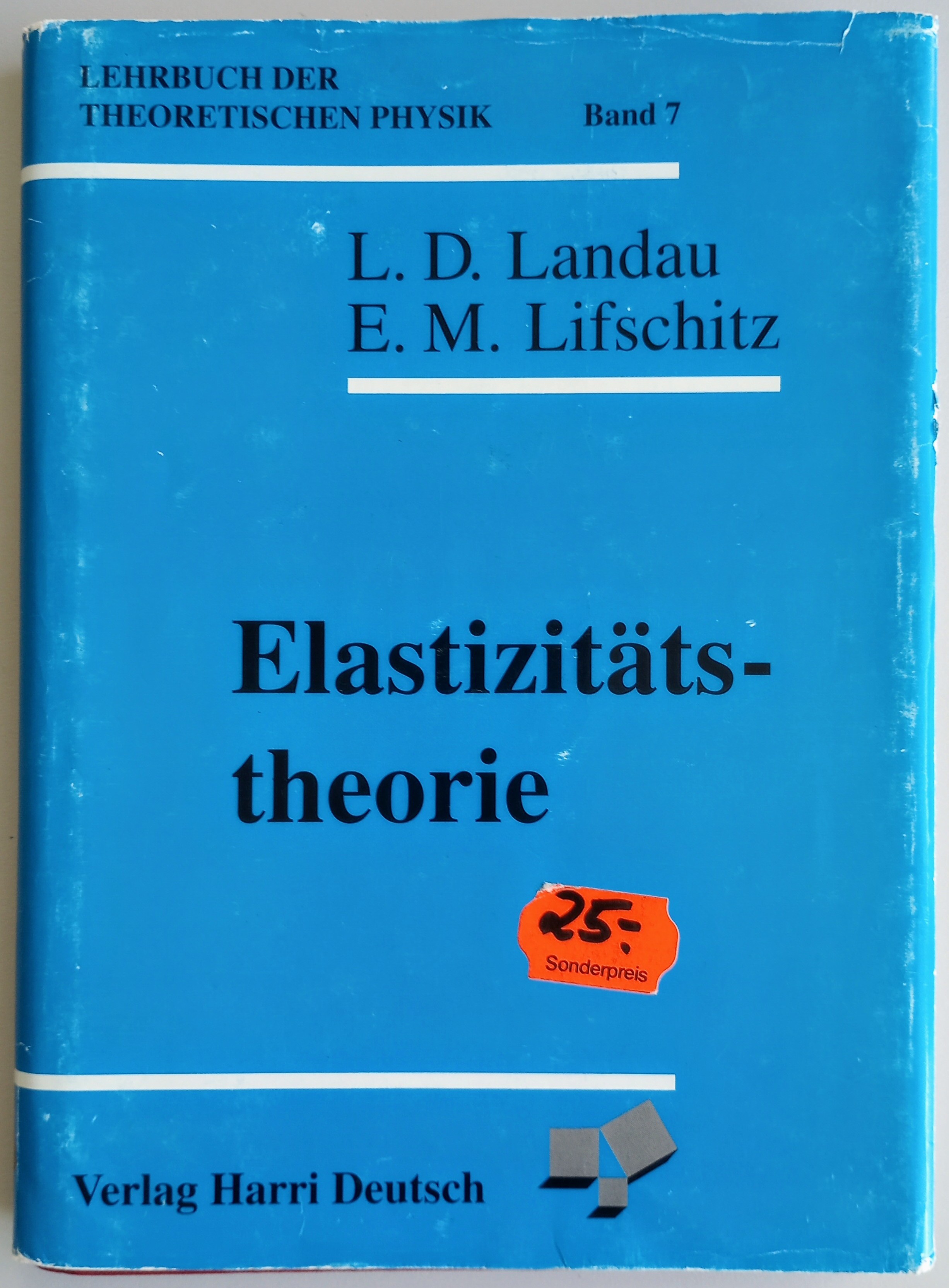}
\hspace{1cm}
(b)
\includegraphics[height=5cm]{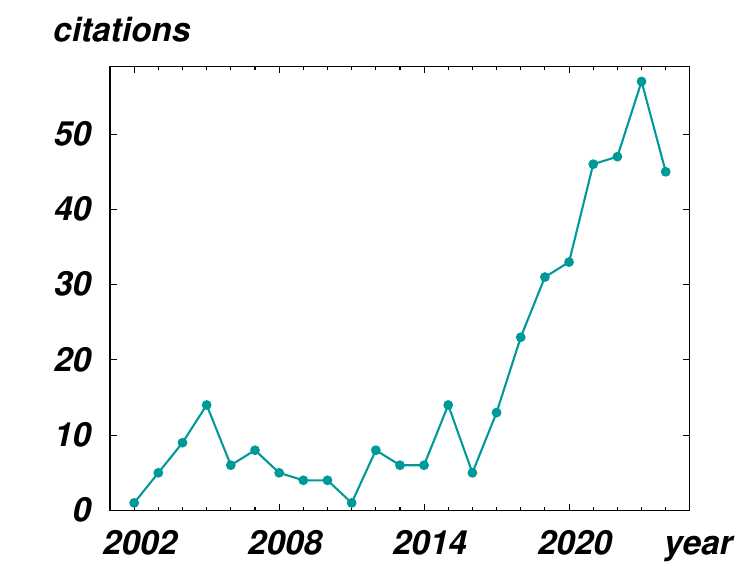}}
\caption{
    (a) Copy of Landau \& Lifshitz's volume 7, Elasticity 
    Theory (in German)~\cite{LLv7} acquired in the summer 2001 
    by one of the authors (P.S.) (special offer for DM$\,$25). \\
    (b) Citation count of Maxim's seminal article 
    \cite{Polyakov:2002yz} on the pressure interpretation 
    of the EMT form factor $D(t)$ since its appearance in 
    October 2002 until end of 2024.
    }
\label{Fig01-LLv7-citations}
\end{figure}

\section{Personal Memories}
\label{Sec-2:memories}

\textit{Some personal memories of P.S.:}
In summer 2001, I acquired in the
university book shop in Bochum the German translation of 
Landau \& Lifshitz Volume 7 on elasticity theory. I was 
happy to get it as a special offer (for DM$\,$25, 
$50\,\%$ discounted, see Fig.~\ref{Fig01-LLv7-citations}a;
the Euro as practical currency was introduced on 1/1/2002).
As I was entering the physics building in Bochum, I met Maxim
who immediately noticed my new book and glanced at it with 
astonishment. I never asked why he was astonished, 
but in the retrospective it became clear. At that time, Maxim
must have been working on the pressure interpretation. It was
Maxim's style to think deeply about the problems he was 
working on. His thoughts were often months and sometimes years 
ahead of his publishing. In this case, we know that the 
seminal work of Maxim about the mechanical interpretation of 
$D(t)$ appeared more than a year later in October 2002~\cite{Polyakov:2002yz} 
(see the citation count from Inspires in Fig.~\ref{Fig01-LLv7-citations}b;
this work was preceded by a short preview of Maxim's main ideas in an 
earlier unpublished preprint in July 2002~\cite{Polyakov:2002wz}).

I have known Maxim since I started in 1996 my Diploma project 
(which later grew into a PhD project) in the Institute for Theoretical 
Physcs II in Bochum under Klaus Goeke who had a unique skill to
create a vibrant and motivating environment where students and 
postdocs could thrive~\cite{Klaus-Festschrift}.  
Maxim was always approachable and happy to discuss with students.
He always stressed that he needed interaction with students to
get new ideas. 
Shortly after the episode with Landau \& Lifshitz Volume 7, I became
postdoc at the Universit\`a degli Studi di Pavia in Italy and interested 
in generalized parton distribution functions (GPDs).
My first self-chosen project was the proof of polynomiality of GPDs 
in the chiral quark-soliton model~\cite{Schweitzer:2002nm} where GPDs and the 
$D$-term were previously computed by Maxim and 
collaborators~\cite{Petrov:1998kf,Penttinen:1999th,Kivel:2000fg}
(along with many other properties).
As it played out, I returned to Bochum two years later, while Maxim 
accepted a professor position in Li\`ege which he held for two years 
before returning to Bochum as a professor. In the subsequent period 
2005-2008, Maxim stimulated and guided the studies in soliton models in 
Bochum~\cite{Goeke:2007fp,Goeke:2007fq,Cebulla:2007ei}, where the 
first insights were obtained on how the pressure and other EMT densities 
may look like in the nucleon. In this period, the Landau \& Lifshitz Volume 7, 
Fig.~\ref{Fig01-LLv7-citations}a, became very helpful for me. But more 
important than that were the discussions and collaborations with Maxim.
Maxim became for me over the years a collaborator, colleague, mentor and friend 
for whom I had always the deepest admiration and respect as a scientist and person.

\textit{Some personal memories of C.L.:}
While I was a graduate student, I met Maxim in 2003 when he had just been appointed Professor at the University of Li\`ege, Belgium. He was very impressive and  at the same time very accessible. It was an intense period where the existence of the strange pentaquark $\Theta^+$, predicted in 1997 by Diakonov, Petrov and Polyakov~\cite{Diakonov:1997mm}, was heavily debated. In 2004, I started a PhD on the chiral quark-soliton model under Maxim's guidance. The following year, I followed him to Bochum where I met so many great and inspiring physicists, including both Victor Petrov and Dmitri Diakonov during one of their visits to Maxim. 

As far as I know, I have been one of Maxim's first PhD students. He told me he considered himself ``too old'' (he was actually only about 40 years old at that time) and wanted a fresh, independent, and unbiased look on the task he assigned me. As a result, I focused essentially on topics closely related to my PhD project. I had no idea about his works on the interpretation of the EMT form factors~\cite{Polyakov:2002wz,Polyakov:2002yz}, and I did not know that he was investigating further the EMT properties within soliton models~\cite{Goeke:2007fp,Goeke:2007fq} while I was in Bochum.

After having worked on the nucleon spin puzzle and charge distributions during my postdoctoral stay in Mainz, I naturally became interested in the EMT and to my great surprise discovered Maxim's seminal work on pressure~\cite{Polyakov:2002yz}. As usual, Maxim's paper was clear and filled with deep physical insights. However, it did not seem to attract much attention in 2012, as indicated by the relatively small number of citations around that period, see Fig.~\ref{Fig01-LLv7-citations} (b). It took about six years to mature some ideas on the extension of Maxim's work to the infinite-momentum frame picture. While we were writing the manuscript in 2018, the first experimental extraction of the quark contribution to pressure inside the proton appeared in Nature~\cite{Burkert:2018bqq}, triggering e-mail exchanges between some theorists. Shortly after, Maxim and Peter uploaded on the arXiv a key review paper on the mechanical properties of hadrons~\cite{Polyakov:2018zvc}. These two articles came out as complete surprises to me. I reached out to Maxim to let him know about our work and we decided to meet a month later to exchange our ideas. It was such a pleasure to be back in Bochum after a long time. My collaborator, Arek Trawi\'nski, presented our results in a seminar and we discussed with Maxim for a couple of days. It was both strange and amusing to realize that, after more than a decade, we independently converged on similar ideas and physics interests. Our paper finally appeared on the arXiv four months later~\cite{Lorce:2018egm}. 

Besides being a great physicist with strong impact in QCD, Maxim was also humble, kind, supportive, and a free-thinker. He remains a model to me and I strive to pass over his legacy to the future generations. Using Maxim's own wording, meeting him has been ``my greatest pleasure''.

\section{The $D$-term, pressure and shear forces}
\label{Sec-3:interpret}

The gravitational form factors (GFFs) were introduced in \cite{Kobzarev:1962wt,Pagels:1966zza}.
In~\cite{Ji:1996ek} Ji noted that the second Mellin moments of GPDs give access to GFFs. The latter are Lorentz-invariant functions that parametrize the EMT matrix elements, just like electromagnetic form factors parametrize the matrix elements of the electromagnetic current. For a nucleon, the symmetric EMT matrix elements can be written as~\cite{Kobzarev:1962wt,Pagels:1966zza,Ji:1996ek}, for a recent review see \cite{Burkert:2023wzr},
\begin{equation}
\begin{aligned}
    \langle p',\vec s\,'|T^{\mu\nu}_a(0)|p,\vec s\rangle&=\overline u(p',\vec s\,')\left[\frac{P^\mu P^\nu}{M}\,A_a(t)+\frac{P^{\{\mu}i\sigma^{\nu\}\lambda}\Delta_\lambda}{M}\,J_a(t)\right.\\
    &\left.\quad+\frac{\Delta^\mu\Delta^\nu-g^{\mu\nu}\Delta^2}{4M}\,D_a(t)+Mg^{\mu\nu}\bar C_a(t)\right] u(p,\vec s),
\end{aligned}
\end{equation}
where $M$ is the nucleon mass, $P=(p'+p)/2$ is the average four-momentum, $\Delta=p'-p$ is the four-momentum transfer, and $t=\Delta^2<0$. The label $a=q,g$ stands for the quark or gluon contribution. We also used for convenience the notation $x^{\{\mu}y^{\nu\}}=(x^\mu y^\nu+x^\nu y^\mu)/2$. The form factor $\bar C_a(t)$ accounts for non-conservation of the separate quark and gluon parts of the EMT, and vanishes identically when summing over all the constituents.

In the forward limit, the $A_a(0)$ represent the fractions of linear momentum carried by quarks or gluons~\cite{Gross:1974cs,Politzer:1974sm} while the $J_a(0)$ describe the corresponding fractions of total angular momentum as shown by Ji \cite{Ji:1996ek}. 
Inspired by this and in analogy to works in gravity\footnote{A reference to the works~\cite{Pomeransky:2000pb,Donoghue:2001qc} can be found above Eq.~(15) of~\cite{Polyakov:2002wz} in which a not yet fully matured preview of Maxim's EMT interpretation ideas is included.}, Maxim defined the static EMT of the nucleon via a Fourier transform in the Breit frame $\vec p\,'=-\vec p=\vec \Delta/2$~\cite{Polyakov:2002yz}
\begin{equation}\label{BFdef}
    \mathcal T^{\mu\nu}_a(\vec r,\vec s)=\int\frac{\ud^3\Delta}{(2\pi)^3}\,e^{-i\vec\Delta\cdot\vec r}\,\frac{\langle p',\vec s|T^{\mu\nu}_a(0)|p,\vec s\rangle}{2E}
\end{equation}
with $E=p'^0=p^0=\sqrt{M^2+\vec\Delta^2/4}$, similarly to the charge distribution~\cite{Sachs:1962zzc}. The time independence of this EMT follows from the fact that the energy transfer $\Delta^0$, which is Fourier conjugate to the time coordinate, vanishes in the Breit frame. 

The components $\mathcal T^{00}_a(\vec r,\vec s)$ and $\mathcal T^{0k}_a(\vec r,\vec s)$ are naturally interpreted as spatial distributions of energy and momentum carried by quarks or gluons. The purely spatial components $\mathcal T^{ij}_a(\vec r,\vec s)$ are then interpreted as the stress tensor\footnote{\label{Footnote-sign-stress-tensor}The stress tensor is usually defined as 
    $\sigma^{ij} = -T^{ij}$ and represents the \emph{external} forces acting on a volume element~\cite{Landau:1975pou}. Sometimes $T^{ij}$ itself is called the stress tensor~\cite{Soper:1976bb} and represents the \emph{internal} forces inside a volume element.}, 
in the sense that if the nucleon were to be considered as a continuous medium, then $\mathcal T^{ij}_a(\vec r,\vec s)$ would characterize the forces exerted by quarks or gluons in an infinitesimal volume at a position $\vec r$ relative to the center of the nucleon. 

In the following we will focus on force interpretation for the total EMT, $\mathcal T^{ij}(\vec r,\vec s) = \sum_a\mathcal T^{ij}_a(\vec r,\vec s)$. For discussions of forces for the separate quark and gluon subsystems, we refer to~\cite{Lorce:2017xzd,Polyakov:2018exb}.

It turns out that for spin-0 as well as spin-$1/2$ particles, the stress tensor is spin-independent and can be expressed as
\begin{equation}\label{Tij}
    \mathcal T^{ij}(\vec r)=\delta^{ij}\,p(r)+\left(\frac{r^ir^j}{r^2}-\frac{1}{3}\,\delta^{ij}\right)s(r).
\end{equation}
The functions $p(r)$ and $s(r)$ are related to each other by the conservation of the total EMT $\nabla^i\mathcal T^{ij}(\vec r)=0$ which implies
\begin{equation}\label{Young-Laplace}
    \frac{\ud }{\ud r}\left(p(r)+\frac{2}{3}\,s(r)\right)=-\frac{2s(r)}{r}.
\end{equation}
The structure of the stress tensor~\eqref{Tij} is the same as that of an anisotropic fluid~\cite{Ruderman:1972aj,Canuto:1974ft,Bayin:1982vw}, where the monopole function $p(r)$ is interpreted as the \emph{isotropic pressure} and the quadrupole function $s(r)$ as the \emph{pressure anisotropy}. Eq.~\eqref{Young-Laplace} can then be understood as the differential form of the Young-Laplace equation, where $s(r)$ can be regarded as the radial distribution of surface tension~\cite{Kirkwood:1949,Marchand:2011}. 

The function $s(r)$ describes also the off-diagonal part of the symmetric stress tensor, and so is related to the distribution of shear forces inside the nucleon. A measure of these shear forces is provided by the so-called $D$-term which is given  by~\cite{Polyakov:2002yz,Polyakov:2018zvc}
\begin{alignat}{7}
    \label{Eq:D-from-s}
       D  &=& \ -\frac{2}{5}\,M
          & \int\ud^3r\,\mathcal T^{ij}(\vec r)\left(\frac{r^ir^j}{r^2}
           -\frac{1}{3}\,\delta^{ij}\right) \ 
          &=&& \ -\frac{4}{15}\,M
          & \int\ud^3r\,r^2\,s(r) , \\
    \label{Eq:D-from-p}
          &=& \frac13\,M
          &\int\ud^3r\,\mathcal T^{ij}(\vec r)\,\delta^{ij}
          &=&& M
          & \int\ud^3r\,r^2\,p(r) ,         
\end{alignat}
provided that the integrals converge. The second representation for $D$ in Eq.~(\ref{Eq:D-from-p}) follows from Eq.~\eqref{Young-Laplace}. 

\section{Concerns in the literature regarding the pressure interpretation}
\label{Sec-4:criticism} 

In the following we list the concerns raised in~\cite{Jaffe:2020ebz,Freese:2021czn,Ji:2021mtz,Freese:2024rkr,Ji:2021mfb,Ji:2022exr,Fujita:2022jus,Czarnecki:2023yqd}, 
which are centered around the following points 
(the sections where we discuss these points
and indicate potential resolutions are mentioned in brackets):
\begin{itemize}
\item 
    The notion of EMT distributions in the Breit frame {is known to be plagued by relativistic recoil corrections, see e.g.~}Refs.~\cite{Jaffe:2020ebz,Freese:2021czn} for recent criticism.
    (Sec.~\ref{Sec-5:comments-Breit-frame})
    
\item
    In ``standard thermodynamics'' the pressure is always positive, while 
    $p(r)$ from the interpretation of $D(t)$ can be negative~\cite{Ji:2022exr}.
    (Secs.~\ref{Sec-6:Poincare-stresses},~\ref{Sec-7:gas+liquid},~\ref{Sec-8:bag-model})

\item 
    In ``fluid dynamics'', an ``effective description in terms of pressure 
    and internal energy [requires] that the particle’s mean free path must 
    be much smaller'' than the size of the system which is the case in QCD 
    ``only at high-temperature and density''~\cite{Ji:2021mtz}.  
    (Sec.~\ref{Sec-9:continuum-description})

\item
    It was argued
    that the $D$-term is negative based on mechanical stability arguments~\cite{Perevalova:2016dln} in agreement with results from hadronic models, lattice QCD, and dispersion relation studies~\cite{Burkert:2023wzr}, but no quantum field theoretical proof of this conjecture exists in QCD~\cite{Fujita:2022jus}. 
    (Sec.~\ref{Sec-9:D-conjecture})
    
\item
    More precisely, $D<0$ appears in systems governed by short-range forces which includes hadrons when QED is neglected, a common approximation in hadronic physics. When the long-range QED effects are present, $D(t)\propto \frac{1}{\sqrt{-t}}$ for $t\to0$ and the $D$-term becomes undefined.\footnote{
        This point is not a criticism of the pressure interpretation. But it is a very interesting point whose resolution may shed light on the next point related to $D$-terms of atoms.} 
    (Sec.~\ref{Sec-10:em-effects}) 
    
\item
    While hadronic $D$-terms are negative (when QED effects are neglected), it was shown that the $D$-term of the hydrogen atom is positive~\cite{Ji:2021mfb,Ji:2022exr,Czarnecki:2023yqd,Freese:2024rkr}. Does this invalidate the pressure interpretation of hadrons? (Sec.~\ref{Sec-11:D-in-atoms}) 
    
\end{itemize}
In the following we explain and review the points of concern in detail, and present 
ideas  and indicate paths along which one may seek their resolutions.

\newpage
\section{Is the use of the Breit frame problematic?}
\label{Sec-5:comments-Breit-frame}

The short answer is: not necessarily. 
Before we go into more detail, it is worth stressing that this issue is not specific to the pressure interpretation. 
Criticism of the interpretation of the nucleon electromagnetic form factors in terms of 3D spatial distributions of electric charge or magnetization~\cite{Sachs:1962zzc} is as old as the interpretation itself~\cite{Yennie:1957skg,Breit:1964ga}. However, this did not prevent Hofstadter from getting the 1961 Nobel Prize for investigating how the electric charge is distributed in nuclei and measuring the proton radius.\footnote{Today, many think first in terms of form factors which 
need to be interpreted. At Hofstadter's time~\cite{Hofstadter:1956qs} the natural approach was the opposite: electric form factors were {\it derived} for (infinitely heavy) nuclei with finite-size electric charge distributions $\rho(r)$ in non-relativistic quantum mechanics in the Born approximation~\cite{Guth,Rose:1948zz,Schiff:1953yzz}. With electron beam energies of up to 150 MeV in the early 1950s~\cite{Schiff:1953yzz}, the assumption of infinitely heavy nuclei and neglect of recoil and spin effects were well justified. Sachs was, to the best of our knowledge, one the first to pursue the nowadays familiar route starting from a fully relativistic setting with electric and magnetic nucleon form factors and introducing their interpretation in some definite frame~\cite{Ernst:1960zza,Sachs:1962zzc}.}

The discussions about the interpretation of form factors continued to this day~\cite{Fleming:1974af,Soper:1976jc,Burkardt:2000za,Kelly:2002if,Belitsky:2003nz,Miller:2010nz,Lorce:2018egm,Lorce:2020onh,Jaffe:2020ebz,Freese:2021mzg,Epelbaum:2022fjc,Panteleeva:2022khw,Panteleeva:2022uii,Panteleeva:2023evj,Chen:2022smg,Li:2022hyf}.
A modern view is that the Breit frame definition can be fully justified by adopting a phase-space perspective, and contenting ourselves with a quasi-probabilistic interpretation~\cite{Lorce:2018egm,Lorce:2020onh}. Alternatively, a pragmatic solution is to consider that the target is so heavy that recoil corrections are negligible. This step is well-justified for (not too light) nuclei. A theoretically rigorous justification of a 3D density interpretation in the Breit frame in the case of the nucleon consists in working in the limit of a large number of colors $N_c$~\cite{Goeke:2007fp,Polyakov:2018zvc,Lorce:2022cle}.

{To avoid recoil corrections while preserving at the same time a probabilistic interpretation, others have proposed to define instead 2D densities in the infinite-momentum frame or using the light-front formalism~\cite{Fleming:1974af,Soper:1976jc,Burkardt:2000za,Miller:2010nz,Lorce:2018egm,Freese:2021mzg,Freese:2022yur}. Averaging over all the directions, one can then define 3D densities in the so-called zero average momentum frame~\cite{Epelbaum:2022fjc,Panteleeva:2022uii}. An interpolation between the Breit frame and the infinite-momentum frame pictures is possible within the phase-space approach, and demonstrates the key role played by Wigner rotations~\cite{Lorce:2018egm,Lorce:2020onh,Chen:2022smg}.}

In what follows, we will use the 3D interpretation in the Breit frame.

\newpage
\section{Poincar\'e stresses and von Laue condition}
\label{Sec-6:Poincare-stresses}

In this section we {show} that {the function} $p(r)$ {obtained} from the interpretation of $D(t)$ 
not only can, but even must be negative in some region of $r$. 
To understand this, it is instructive to review two (not unrelated) 
aspects of the early history of particle physics and special relativity. 

First, the discovery of the electron by J.~J.~Thompson in 1897 triggered efforts 
to develop classical models of the electron, which was an endeavour at the 
center of attention of the founders of special relativity~\cite{Bialynicki-Birula:1983ace},
with works by Thomson, Abraham, Lorentz, Poincar\'e, Einstein, Wien, Planck, 
Sommerfeld, Langevin, Ehrenfest, Born, Pauli, von Laue and others~\cite{Miller}. 
In these early pre-quantum days, the electron was thought to be a little sphere 
carrying some classical electric charge distribution, and conflicting results 
were obtained like  $E_e = \frac43\,m_ec^2$ for the rest energy of the electron.
Poincaré recognized the necessity to introduce forces of 
non-electromagnetic origin,\footnote{Strictly speaking, this resolves 
    some but not all inconsistencies in classical calculations of the 
    electromagnetic mass of a particle. For a pedagogical exposition 
    see~\cite{Griffiths}.} 
known as ``Poincaré stresses'', to compensate the Coulomb repulsion 
of the electron's own charge distribution and ensure a bound system~\cite{Poincare}.\footnote{Even after
    the pointlike nature of the electron became clear, with its classical 
    divergent electrostatic self-energy put routinely under the carpet 
    in dimensional regularization~\cite{Czarnecki:2023yqd},  
    the formulation of consistent classical models of finite-size 
    charged particles remained an intellectually challenging problem 
    with interest until the present~\cite{Dirac:1962iy,Boyer:1982ns,Schwinger:1983nt,Bialynicki-Birula:1993shm,Rohrlich}. \label{Footnote-classical-models}}

Second, it is well known that Einstein's original proof of $E=mc^2$ \cite{Einstein} and 
subsequent variants thereof were based on simplifying assumptions
like low speeds $v\ll c$, and cannot be considered valid
``fully relativistic'' proofs, as Einstein did 
not yet have the concept of a conserved EMT $T^{\mu\nu}$. 
The generally accepted proof was given by Max von Laue in 1911
for closed systems with a time-independent EMT~\cite{Laue:1911lrk},
and generalized by Felix Klein in 1918 to closed systems with an 
arbitrary time dependence \cite{Klein}, for reviews of the interesting
history see~\cite{Ohanian}. Maxim's static EMT in Eq.~\eqref{Tij}
corresponds to a closed, time-independent system in von Laue's proof
of $E=mc^2$, for which an essential ingredient is the so-called
von Laue condition~\cite{Laue:1911lrk}  
\be\label{Eq:von-Laue}
    \int \ud^3r\;T^{ii}(r) = 0 \quad \Leftrightarrow \quad
    \int\limits_0^\infty \ud r\;r^2p(r) = 0 
\ee
that follows from EMT conservation. It is a necessary (but not sufficient\footnote{EMT conservation implies that momentum current densities $T^{ij}(\vec r)$ are divergenceless. This means only that the system is at equilibrium and says nothing about its stability, contrary to what is suggested in~\cite{Ji:2021mfb}.}) 
condition for stability of a system, and can be understood as a 
field-theoretic version of the virial theorem~\cite{Lorce:2021xku}. 

After these preparations, it is clear why $p(r)$ in Eq.~\eqref{Tij}
cannot be positive everywhere: it must have at least one node and change sign 
somewhere to comply with the von Laue condition \eqref{Eq:von-Laue}. 

\section{Gases, liquids, positive and negative pressures}
\label{Sec-7:gas+liquid}

The above discussion of Poincar\'e stresses indicates that we cannot simply rely on the
intuition about pressure from ``standard thermodynamics''~\cite{Ji:2022exr} 
familiar, e.g., from the ideal gas equation of state $PV=NkT$, 
where $P$ is always positive. The point is that
an ideal gas is an inherently unstable system: without a container, 
i.e.\ a fixed volume $V$, the ideal gas would immediately expand
and disperse. Providing a container corresponds to introducing
cohesive forces, the Poincar\'e stresses, necessary to form 
a stable, closed system.
(Notice that in thermodynamic equilibrium and when external forces are absent, both the 
density $\rho =N/V$ and pressure $P$ are constant throughout 
the volume occupied by the gas. In presence of external forces like gravity, the density and pressure are not constant anymore and containers are not 
always necessary, as illustrated e.g.~by earth atmosphere, stars and confined 
gas clouds in galaxies.)

An actual gas is more realistically described by the van der Waals equation of state
$(P+\frac{aN^2}{V^2})(V-Nb)=NkT$, where $a>0$ accounts for the attractive 
forces between the particles and $b>0$ corrects for their finite volume. 
Rewriting this equation as $P=\frac{NkT}{V-Nb}-\frac{aN^2}{V^2}$, we see that 
attractive forces between the particles reduce the pressure, but of
course not enough to reverse the sign: the attractive forces in a gas are relatively small 
compared to the inertial forces and so the total pressure remains positive. 
A van der der Waals gas still needs a container to form a closed system.

Loosely speaking, the region with $p(r)>0$ corresponds to where 
the matter in the system sits. This matter would disperse without 
the Poincar\'e stresses in the region with $p(r)<0$ which bind the system. 
Although this is a gross oversimplification in general, in one ideal system
this is literally the case, namely in the liquid drop which was, 
in fact, explored by Maxim as a pedagogical illustration for his 
newly introduced concepts in Ref.~\cite{Polyakov:2002yz}.

The liquid drop is bound by surface tension (one may think of
astronauts and cosmonauts playing with  water drops 
in space craft in Earth's~orbit).
The surface tension $\gamma$ enters the shear force as 
$s(r)=\gamma\,\delta(r-R)$ where $R$ is the radius of the drop.
Since $s(r)=0$ for $r\neq R$, Eq.~\eqref{Young-Laplace} dictates
that $p(r)$ is constant for $r\neq R$. More precisely, $p(r)$ is 
equal to the constant hydrostatic pressure $p_0$ for $r<R$ where 
the liquid is, and zero in the region $r>R$ where no
liquid is present. Solving the Young-Laplace 
equation with these boundary conditions for given 
$s(r)=\gamma\,\delta(r-R)$ yields for the pressure
$p(r)=p_0\Theta(R-r)-\frac23\,\gamma\,\delta(R-r)$,
and the von Laue condition \eqref{Eq:von-Laue} 
yields the Kelvin relation $p_0 = 2\gamma/R$ for the pressure 
inside a drop in terms of $\gamma$ and $R$~\cite{Kelvin}. 
We see that, for an ideal drop, it is $p(r)<0$ only at one 
single point, namely at $r=R$ where the surface
tension in $s(r)=\gamma\,\delta(r-R)$ acts, providing in this
way the Poincar\'e stresses necessary to bind the system.

Since many bulk properties of nuclei can be approximately
described in the liquid-drop model, Maxim made predictions
for the dependence of nuclear $D$-terms on the mass number $A$
as $D\propto -\,A^{7/3}$ \cite{Polyakov:2002yz} which were confirmed 
in more elaborate nuclear models, like Walecka \cite{Guzey:2005ba} 
or Skyrme model~\cite{GarciaMartin-Caro:2023toa}, 
and await experimental tests.

Although this is not always discussed in standard thermodynamics textbooks,
negative pressures do exist in thermodynamics.
From a microscopic perspective, the ``standard thermodynamic'' pressure is understood as the macroscopic manifestation of a large collection of particles zipping around and bouncing off walls. As discussed above, in classical gases the kinetic energy dominates and pressure is naturally positive. 

In liquids, where the cohesive forces are stronger than in gases, it is usually taught that the hydrostatic pressure is isotropic and positive. This is true in most practical cases since the density usually does not change much in the bulk of a liquid. But the situation is different at the interfaces, where the density gradients become large and surface tension effects (i.e.~anisotropic stresses) have to be invoked to describe the observations. Bulk pressure can even become negative when the liquid is in a metastable state, e.g.~when it is trapped in a closed system and cannot expand or contract freely~\cite{Mercury,Debenedetti}. In particular, negative pressure resulting from evaporation provides a simple explanation for why water can rise well above 10 m in the trees~\cite{Wheeler,Debenedetti}, a phenomenon that cannot be quantitatively explained with capillarity arguments.

For inhomogeneous media at rest the pressure, defined as the magnitude of the force per unit area normal to the surface element, is not isotropic and receives a contribution from both $p(r)$ and $s(r)$. 
In the context of neutron stars, the combinations $p_r(r)=p(r)+\frac{2}{3}\,s(r)$ and $p_t(r)=p(r)-\frac{1}{3}\,s(r)$ are respectively called \emph{radial pressure} and \emph{tangential pressure}~\cite{Bayin:1982vw}, and correspond to the normal stresses along the radial and tangential directions. For static fluid spheres, it is usually expected that $p_r(r)\geq 0$~\cite{Herrera:1997plx,Mak:2001eb}. In the familiar non-relativistic situations, we typically observe that $p(r)\gg s(r)$ (except at the interfaces) and so the positivity of the radial pressure effectively reduces to the positivity of the isotropic pressure.

In summary, pressure is not always isotropic and can even become negative. Therefore, the fact that the function $p(r)$ does necessarily become negative in some region does not jeopardize its interpretation as isotropic pressure, but simply indicates that there are situations where the isotropic part of the attractive or confining forces dominates.

\newpage

\section{Poincar\'e stresses in bag model}
\label{Sec-8:bag-model}

It is instructive to discuss the Poincar\'e stresses  
in the bag model~\cite{Chodos:1974je}. For pedagogical purposes let us start 
with its predecessor, the Bogoliubov model~\cite{Bogoliubov} (the history of 
the two models is reviewed in~\cite{Thomas+Weise}, Sec.~8.4).
The Bogoliubov model does not describe a stable nucleon.
It corresponds basically to the bag model with $N_c$ 
non-interacting massless quarks in anti-symmetric color state, 
inserted in a cavity of radius $R$ with a boundary condition 
such that there is no energy-momentum flow across the surface 
(this is how quarks are confined). The only difference with the
bag model is that there is no bag contribution in the 
Bogoliubov model. The EMT in the Bogoliubov model is given
solely in terms of $T^{\mu\nu}_{\rm Dirac}$ of the quarks.
In particular, the nucleon mass is given by 
$M_{\rm Bogo}=N_c\,\omega_0/R$, where $\omega_0 \simeq 2.04$
is the (ground state) solution of a transcendental 
equation dictated by the boundary condition. 
The free parameter $R$ must be fixed by 
hand to reproduce the experimental value of the nucleon mass.
The pressure in the Bogoliubov model is $p(r)_{\rm Bogo}>0$~\cite{Neubelt:2019sou},
i.e.\ according to the von Laue condition \eqref{Eq:von-Laue}
we deal with an unstable system: the nucleon explodes.
This can be seen equivalently from 
$M_{\rm Bogo}=N_c\omega_0/R$, namely: 
if one does not fix the bag radius by hand, then 
the nucleon mass is minimized for $R\to\infty$.
The reason for this obvious: the model lacks the Poincar\'e
stresses and the quarks disperse due to the positive 
Fermi pressure. (At this point it is helpful to consider
the large-$N_c$ limit, so that we deal with a ``macroscopic number''
of quarks and the notion of thermodynamic pressure can be defined.)

This problem is remedied in the bag model by supplementing 
$T^{\mu\nu}_{\rm Dirac}$ in the EMT with the bag contribution\footnote{In cosmology, 
    dark energy is often modeled as a bag-like contribution with $R$ equal to the size of the Universe and $B$ proportional to the cosmological constant~\cite{Frieman:2008sn,Martin:2012bt}. This dark energy provides a negative pressure contribution which drives the observed accelerating expansion of the Universe. The analogy between the cosmological constant and pressure contributions in hadrons has been discussed in~\cite{Teryaev:2013qba,Teryaev:2016edw,Liu:2023cse}.}
$T^{\mu\nu}_{\rm bag} = B\,\Theta(R-r)\,g^{\mu\nu}$,
where $B>0$ is the bag constant. 
This has two effects: (i) the mass in the bag model
is given by $M_{\rm bag}=N_c\,\omega_0/R+\frac43\pi R^3B$ and 
(ii) the pressure receives a negative constant contribution $(-B)$.
One of the two model parameters, $R$ and $B$, can be fixed by 
minimizing $M_{\rm bag}$ with respect to $R$ which yields\footnote{
    Evaluating the nucleon mass at the minimum gives 
    $M_{\rm bag} = \frac43N_c\omega_0/R$. It is amusing to note
    that $M_{\rm bag} = \frac43\,M_{\rm Bogo}$, which is reminiscent 
    of the famous $\frac43$ paradox in the incomplete classical electron 
    models, see Sec.~\ref{Sec-6:Poincare-stresses}. Notice that here
    the Bogoliubov model is incomplete, lacks Poincar\'e stresses, 
    and {\it misses} $\frac14$ of the nucleon mass.}
the condition $N_c\,\omega_0 = 4\pi R^4B$ sometimes
referred to as virial theorem in the bag model. If this condition 
is satisfied, then the von Laue condition \eqref{Eq:von-Laue} is
also satisfied. The equivalence of the von Laue condition and the 
virial theorem has been observed in many models, see e.g.~\cite{Goeke:2007fp,Goeke:2007fq,Cebulla:2007ei,Kim:2012ts,
Mai:2012yc,Mai:2012cx,Jung:2013bya,Jung:2014jja}
and discussed in the general field-theoretic context in~\cite{Lorce:2021xku}. 
Thus, the Poincar\'e stresses are explicitly incorporated in
$T^{\mu\nu}_{\rm bag} = B\,\Theta(R-r)\,g^{\mu\nu}$ and ensure in this
way a consistent model of a stable nucleon. 

The bag model provides another instructive lesson, namely 
regarding the form factor $\bar{c}^a(t)$ and nicely illustrating
the usefulness and theoretical consistency of the EMT form factor
interpretation. One of the earliest model studies
of EMT form factors was presented in the bag model~\cite{Ji:1997gm}
where, starting from the well-known quark wave functions, 
the EMT form factors of quarks were computed and found non-zero. 
Interestingly, the EMT form factors 
$A(t)=\sum_q A^q(t)$, $J(t)=\sum_q J^q(t)$, and $D(t)=\sum_q D^q(t)$ 
(summed over quark flavors) are given entirely in terms of quarks only. 
This can be seen, e.g., from the fact that 
$\sum_q A^q(0)=1$ and $\sum_q J^q(0)=\frac12$~\cite{Ji:1997gm}.

However, the situation is different for the fourth EMT form factor ${\bar C}^a(t)$ since one finds that 
$\sum_q {\bar C}^q(t) \neq 0$~\cite{Ji:1997gm}. Clearly something is missing:
we should have $\sum_a {\bar C}^a(t)= 0$ when summing over {\it all} 
degrees of freedom because the EMT is certainly conserved in this model~\cite{Chodos:1974je} and the von Laue condition \eqref{Eq:von-Laue} is 
satisfied~\cite{Neubelt:2019sou}. 
What we are missing is the contribution from the bag, but we encounter here an unpleasant problem: 
it is not possible to treat the bag as a degree of freedom,
determine its wave function, and compute in this way 
${\bar C}^{\rm bag}(t)$~\cite{Ji:1997gm}. 

Fortunately, at this point the 3D interpretation of EMT 
densities comes to the rescue. The bag contribution 
$T^{\mu\nu}_{\rm bag} = B\,\Theta(R-r)\,g^{\mu\nu}$ to the EMT can be Fourier 
transformed and yields exactly 
${\bar C}^{\rm bag}(t)=-\sum_q {\bar c}^q(t)$~\cite{Neubelt:2019sou}. 
This corresponds to {\it inverting} the EMT form factor interpretation
and is, to the best of our knowledge, the only way
to compute ${\bar C}^{\rm bag}(t)$ in this model. This 
provides an independent proof of EMT conservation in the bag model
and illustrates the theoretical consistency of the EMT distribution 
formalism. Notice that the calculation in Ref.~\cite{Neubelt:2019sou} 
was carried out in the large-$N_c$ limit, where the 3D interpretation 
of EMT form factors is justified~\cite{Goeke:2007fp,Polyakov:2018zvc}.

\newpage
\section{The mean free path and continuum description}
\label{Sec-9:continuum-description}

In classical physics, matter is described at the microscopic level as a large collection of particles that interact with each other. At the macroscopic level, one can use instead an effective continuum description. This is typically justified when the size of the volume over which one averages the microscopic details is much larger than the typical mean free path of the particles. In quantum theory, the particle-like description makes sense provided that the typical sizes of the problem are much larger than the Compton wavelength.\footnote{When the particle is moving, the characteristic wavelength is given in general by the inverse of the energy and not the inverse of the mass~\cite{Burkardt:2000za}.} Below the Compton wavelength, one should rather switch to a wavelike picture.

We naturally tend to think of hadrons as bound states of quarks and gluons. However, it is not clear at all that a particle-like picture is well adapted to describe the structure of light hadrons, since the latter are typically states with an \textit{indefinite} number of quarks and gluons. We should keep in mind that \textit{fields} are the fundamental degrees of freedom in quantum field theory.\footnote{Note, e.g., that the Casimir effect, which consists of a modification of the vacuum expectation value of the electromagnetic field energy, results in a force per unit area acting on two parallel conducting plates, sometimes refered to as the Casimir pressure~\cite{Mostepanenko:1988,Klimchitskaya:2009cw}.} Particles appear only as effective degrees of freedom arising from the complicated collective dynamics and excitations of the fields. 

In the context of hadronic physics, the notion of pressure is defined directly in terms of the fundamental microscopic EMT, and not in terms of an effective macroscopic one. There is therefore no need to justify it by mean free path arguments. This is to be contrasted with studies of the quark-gluon plasma, where a macroscopic effective description of the system, motivated by the large number of quasifree excitations of the quark and gluon fields, is usually adopted~\cite{Shuryak:2014zxa}.

\newpage

\begin{figure}[b!]
\centerline{
\ \hspace{-10mm}
\includegraphics[height=5cm]{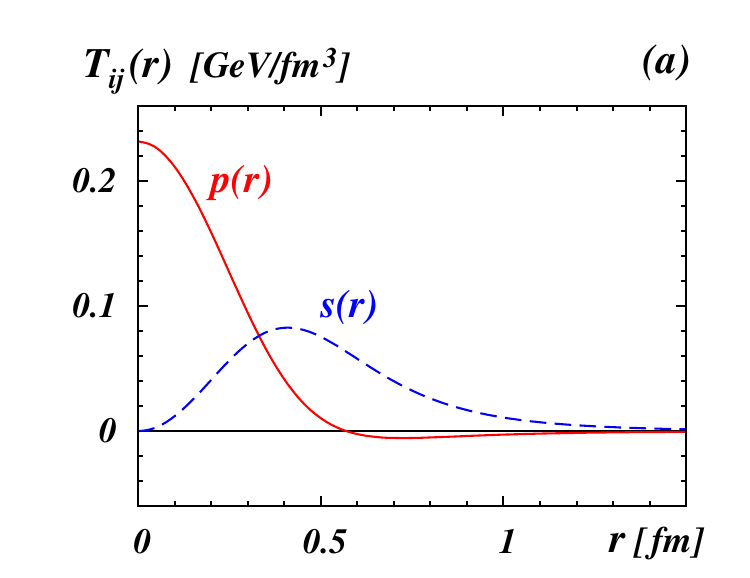}
\includegraphics[height=5cm]{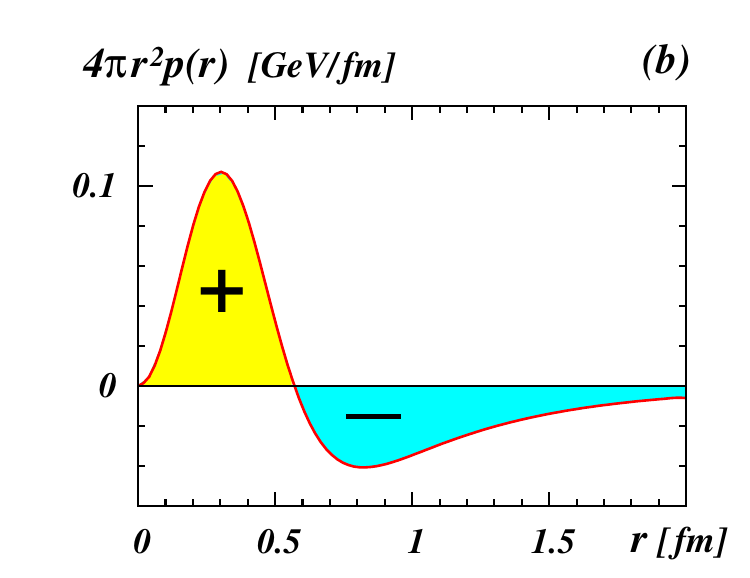}}
\caption{
    (a) Shear and pressure distributions, $s(r)$ and $p(r)$, describing 
    the nucleon stress tensor $T_{ij}(r)$ in Eq.~\eqref{Tij} as 
    obtained from the chiral quark-soliton model~\cite{Goeke:2007fp}.
    (b) $4\pi r^2p(r)$ illustrating how the von Laue condition 
    is satisfied in the said model: the positive and negative
    contributions compensate each other according to
    Eq.~\eqref{Eq:von-Laue}.}
\label{Fig02-CQSM}
\end{figure}

\section{The negative $D$-term sign conjecture}
\label{Sec-9:D-conjecture}

Maxim, in collaboration with Christian Weiss, proved a soft 
pion theorem stating that the quark contribution to the pion 
$D$-term is $D^q =-A^q(0)$ in chiral limit~\cite{Polyakov:1999gs}. 
This holds also for the gluon part implying for the total 
$D$-term of the pion $D_{\rm pion}=-1$. This is true also for other Goldstone 
bosons associated with chiral symmetry breaking, namely kaons and $\eta$-meson, 
and was known from early works on hadronic decays of quarkonia~\cite{Voloshin:1980zf,Leutwyler:1989tn,Voloshin:1982eb,Novikov:1980fa}.
The $D$-term of the nucleon was found negative in the bag model~\cite{Ji:1997gm} and chiral quark-soliton model~\cite{Petrov:1998kf,Kivel:2000fg}. Maxim also predicted that nuclear $D$-terms are negative~\cite{Polyakov:2002yz}. 

The first theoretical study of spatial EMT distribution was, to the best 
of our knowledge, carried out in the chiral quark-soliton model in 
Ref.~\cite{Goeke:2007fp} of which Maxim is a co-author. 
Stressing the relevance of the von Laue condition~\cite{Laue:1911lrk} and its connection to the virial theorem, 
this work among others also raised the suspicion that the negative sign 
of $D$ might be a natural consequence of stability~\cite{Goeke:2007fp},
as illustrated in Fig.~\ref{Fig02-CQSM}.

The shear and pressure distributions $s(r)$ and $p(r)$ from the chiral 
quark-soliton model exhibit a remote similarity to the liquid drop picture,
see Fig.~\ref{Fig02-CQSM}a: throughout $s(r)\ge 0$, while $p(r)>0$ in the 
inner region and~$p(r)<0$ in the outer region, whereby the $\delta$- and 
$\Theta$-functions of the ideal liquid drop, see Sec.~\ref{Sec-7:gas+liquid},
are strongly smeared out in the nucleon. This is not surprizing after all, as the nucleon is considerably more diffuse than a nucleus.

Fig.~\ref{Fig02-CQSM}b shows how the internal forces balance each other 
in the nucleon: $p(r)>0$ in the center region corresponding to repulsive 
forces are balanced by the attractive forces, Poincar\'e stresses, in the 
outer region where $p(r)<0$. The sign change occurs at the node 
$r_0 = 0.57\,{\rm fm}$ (in this model). Multiplying $p(r)$ by $r^2$ yields 
the integrand of the von Laue condition and
is shown in Fig.~\ref{Fig02-CQSM}b.  The areas above and below the $r$-axis 
compensate each other exactly according to 
Eq.~\eqref{Eq:von-Laue}.
(A factor $4\pi$ is included in Fig.~\ref{Fig02-CQSM}b to compare  
the $\mbox{force} = (\mbox{pressure})\times(\mbox{area}\;4\pi r^2)$ 
to the string tension in the confining QCD potential known, e.g., 
from quarkonium studies $V_{\rm conf}(r) = k\,r$ with 
$k\simeq 1\,\rm GeV/fm$~\cite{Barnes:2005pb}. The magnitude of forces 
in the nucleon is understandably nowhere close to the string tension.)

Notice that the von Laue condition \eqref{Eq:von-Laue} would be equally 
well satisfied if the picture in Fig.~\ref{Fig02-CQSM}b had reversed 
signs, with a negative pressure in the inner region and a positive pressure in 
the outer region. From the point of view of mechanical stability, 
however, this would correspond to a seemingly unphysical situation because 
(i)  nothing would prevent the matter within the node $r_0$ 
     from collapsing into the center, and 
(ii) nothing would prevent the matter beyond 
    $r_0$ from dispersing into infinity
(which reflects the fact that the von Laue condition is necessary 
but not sufficient for stability). 

Thus, the sign pattern in Fig.~\ref{Fig02-CQSM}b is a consequence
of mechanical stability (we will elaborate on this point below).
Now, compliance with the von Laue condition \eqref{Eq:von-Laue} 
in conjunction with the physical sign pattern of $p(r)$ dictates 
that the $D$-term $D=4\pi M \int_0^\infty \ud r\,r^4p(r)$ is negative,
since the additional weight of $r^2$ under the $r$-integration as 
compared to Eq.~\eqref{Eq:von-Laue} and Fig.~\ref{Fig02-CQSM}b
suppresses the contribution of the small-$r$ region and enhances
that of the large-$r$ region~\cite{Goeke:2007fp}. 

This physically appealing insight was supported by 
subsequent works where negative $D$-terms were obtained in 
other hadron models, lattice QCD, dispersion relations, 
$Q$-ball systems, and phenomenological hints~\cite{Goeke:2007fq,Cebulla:2007ei,Kim:2012ts,Mai:2012yc,Mai:2012cx,Jung:2013bya,Jung:2014jja,Pasquini:2014vua,Perevalova:2016dln,Hudson:2017oul,Hudson:2017xug,Burkert:2018bqq,Polyakov:2018zvc,Lorce:2018egm,Shanahan:2018nnv,Polyakov:2019lbq,Varma:2020crx,Neubelt:2019sou,Metz:2021lqv,Gegelia:2021wnj,Owa:2021hnj,Pefkou:2021fni,Lorce:2021xku,Freese:2022jlu,Freese:2022ibw,Freese:2022yur,Fu:2022rkn,Lorce:2022cle,Panteleeva:2023aiz,Amor-Quiroz:2023rke,Fu:2023ijy,Wang:2023bjp,Maynard:2024wyi,Wang:2024abv}. 
A physical pressure sign pattern and negative $D$-term
are necessary but not sufficient mechanical stability conditions
and apply equally to unstable systems (e.g.\ Roper resonance, 
$\rho$-meson, $Q$-clouds) which correspond to local minima 
of the action in the pertinent quantum number sector.

The above observations were solidified by 
Irina Perevalova, Maxim  
et al. in~\cite{Perevalova:2016dln}, where 
the conclusion that $D<0$ was drawn based on the assumptions that 
(i) the pressure interpretation is exact, 
(ii) mechanical stability criteria {\sl can} be applied to hadrons,
(iii) the densities $p(r)$ and $s(r)$ decay at long-distances faster
than $\frac1{r^4}$.
The first point is justified for the nucleon in the large-$N_c$ limit. 
The second is an assumption which, if one accepts its validity, can be 
explored as shown below. The third is a technical assumption which is
satisfied for hadrons, provided one neglects electromagnetic effects 
(we will come back to this point below).

The argument is that at any point inside a mechanical stable system 
the radial pressure $p_r(r)$ must satisfy the local stability criterion
\be\label{Eq:p-radial}
    p_r(r)=p(r)+\frac{2}{3}\,s(r) \ge 0, 
\ee
or else one would encounter mechanical instability~\cite{Perevalova:2016dln}.
(This criterion has already been mentioned 
in Sec.~\ref{Sec-7:gas+liquid} in the context 
of neutron stars.) The conclusion follows immediately from the fact 
that if \eqref{Eq:p-radial} holds, then $4\pi\int_0^\infty \ud r\,r^4p_r(r) > 0$ which can be expressed with 
the help of Eqs.~(\ref{Eq:D-from-s},~\ref{Eq:D-from-p}) as 
$(-3D)/(2M)$, q.e.d.~\cite{Perevalova:2016dln}.

To the best of our knowledge, this is presently the most compelling explanation
of why $D<0$ is found in all physically sound (stable, metastable, unstable)
systems governed by short-range forces. In the next section we will
discuss what happens in the presence of long-range forces.

Before going into that, it is worth discussing where the 
Poincar\'e stresses come from in the chiral quark-soliton model.
They are provided by the strong chiral forces in the soliton (mean) 
field. Such a mean field picture is known to arise in QCD in the 
large-$N_c$ limit~\cite{Witten:1979kh}. At long distances the 
distributions behave like $s(r) =  3 \,\frac{a}{r^6}$ and 
$p(r) = -\,\frac{a}{r^6}$ in the chiral limit with 
$a=(\frac{3g_A}{8\pi f_\pi})^2$ for large $N_c$, where $g_A=1.26$ 
and $f_\pi=91\,\rm MeV$ denote, respectively, the nucleon axial 
coupling and pion decay constant 
(off the chiral limit, when the pion mass $m_\pi\neq 0$,
the EMT densities are exponentially suppressed with tails
proportional to $e^{-2m_\pi r}$)~\cite{Goeke:2007fp}.

It is well known that the exchange of spin-0 particles yields 
attractive forces, and the chiral Goldstone bosons are spin zero. 
In this sense, the Poincar\'e stresses in the nucleon emerge from 
the pion cloud (see Ref.~\cite{Meissner:2007tp} for a careful 
review of the history and folklore of this term). 

The results in Fig.~\ref{Fig02-CQSM} are from a specific
chiral model, but the long-distance behavior 
$s(r) =  3 \,\frac{a}{r^6}$ and $p(r) = -\,\frac{a}{r^6}$ 
is model independent and dictated by chiral symmetry. 
This fact was in turn explored by Maxim and Jambul Gegelia
to determine a conservative upper bound for the nucleon $D$-term 
of $D \le -(0.20\pm 0.02)$~\cite{Gegelia:2021wnj}.
Noteworthy, the upper bound indicates that the $D$-term is
negative.
In the derivation of this bound, long-range forces were neglected 
whose impact we shall discuss in the next section.

\newpage
\section{Long-range forces and $D$-terms of charged particles}
\label{Sec-10:em-effects}

The $D$-term is arguably the particle property most sensitive to 
changes in the parameters or dynamics of a theory~\cite{Hudson:2017xug}. 
For instance, masses of nuclei grow linearly 
with mass number $A$ while $D\propto -\,A^{7/3}$~\cite{Polyakov:2002yz}.
Similarly, masses of large $Q$-balls grow approximately linearly with 
their charge $|Q|$, while their $D$-terms grow like $D\propto -\,|Q|^{7/3}$~\cite{Mai:2012yc}.\footnote{The appearance of the same power 7/3 in 
    both systems is not accidental: the ground states of large nuclei 
    and those of $Q$-balls share basic similarities with liquid drops
    (radii growing like $A^{1/3}$ or $|Q|^{1/3}$, one can define a
    surface tension, etc.)~\cite{Coleman:1985ki}. }
Or, if one goes to high-lying excitations in the spectra
of $Q$-balls and baryons (in the bag model), then 
$D\propto -\,M^{8/3}$ as the masses of the excited states increase~\cite{Mai:2012cx,Neubelt:2019sou}.\footnote{Whether  
    the appearance of the same power 8/3 in the case of excitations 
    in these very different systems has deeper underlying reasons or 
    is accidental is unknown to~us.}
However, the strongest sensitivity of the  $D$-term observed so far is
when electromagnetic effects are included~\cite{Varma:2020crx}, which
were neglected in prior hadronic studies 
(with few exceptions~\cite{Kubis:1999db,Donoghue:2001qc}). 

In Ref.~\cite{Varma:2020crx} the {\it classical} model of the 
proton by Bia\l ynicki-Birula~\cite{Bialynicki-Birula:1993shm}
was explored. The latter is remarkable 
(cf.\ footnote~\ref{Footnote-classical-models})
in that it provided the first classical framework with 
Poincar\'e stresses introduced in a dynamical~way 
(as opposed to an adhoc way in prior works).
This is a classical relativistic field theory which describes the 
proton as a static pressureless dust bound within a radius 
$R_{\rm dust}\simeq1\,\rm fm$ (which emerges from the
model dynamics) by the interplay of strong massive 
scalar and vector as well as electro\-static forces. 
The strong forces are taken from nuclear models, i.e.\ the dust 
is bound by residual strong forces~\cite{Bialynicki-Birula:1993shm}. 

An inspection of the contributions to the von Laue integral
\eqref{Eq:von-Laue} from the scalar (S), vector (V) meson
and electrostatic (em) fields leads to 
\begin{alignat}{5}
 & \textstyle{\int_0^\infty} \ud r\,r^2p(r)_S     &=& -10.916\,{\rm MeV}, \quad && \tfrac{g_S^2}{4\pi\hbar c}=7.29, \nonumber\\
 & \textstyle{\int_0^\infty} \ud r\,r^2p(r)_V     &=& +10.891\,{\rm MeV}, \quad && \tfrac{g_V^2}{4\pi\hbar c}=10.8, \quad  &\nonumber\\
 & \textstyle{\int_0^\infty} \ud r\,r^2p(r)_{em}  &=& \;+0.025\,{\rm MeV}, \quad && \tfrac{e^2}{  4\pi\hbar c}=\tfrac1{137}.
 \label{Eq:von-Laue-in-classical-model}
\end{alignat}
The coupling constants included in Eq.~\eqref{Eq:von-Laue-in-classical-model}
show that strong forces are three orders of magnitude stronger than electric 
ones, and the latter make an accordingly minuscule contribution to the
balance of forces.

The negative sign signals attractive (scalar) forces, while the positive 
signs indicate repulsive (vector, electrostatic) forces. The contributions
to $p(r)$ and other EMT distributions from the massive strong fields exhibit
Yukawa-type tails proportional to $e^{-m_i r}$ with $i=S,\,V$. The stronger
repulsive vector fields isare more massive with $m_V = 783\,\rm MeV$ ($\omega$-meson mass),
have a shorter range, and make $p(r)$ positive in the inner region.  
The somewhat weaker and lighter scalar fields with $m_S = 550\,\rm MeV$ (``$\omega$-meson'' mass) have a longer range, dominate the outer region, and
provide there the Poincar\'e stresses. The electromagnetic effects in 
Eq.~\eqref{Eq:von-Laue-in-classical-model} seem to play a naturally subordinate 
role for the structure of the proton. The results of the classical 
proton model look qualitatively very similar to those in Fig.~\ref{Fig02-CQSM}, 
except that $s(r)$ and $p(r)$ are about one order of magnitude smaller 
(since the residual nuclear forces of the classical model are significantly
weaker than the strong mean field forces in the chiral quark-soliton model).
The electromagnetic effects make an impact only at long distances, beyond the 
range displayed in Fig.~\ref{Fig02-CQSM}. 

The strong fields govern the structure of the inner region in the classical
proton model and somewhat beyond  until about $r\simeq 2\,\rm fm$. But they 
are exponentially suppressed, and around $r\gtrsim\mbox{(2-3)}\,\rm fm$ 
the electrostatic contribution becomes comparably strong, and makes $s(r)$ 
and $p(r)$ change sign again. The Maxwell stress tensor 
$T^{ij} = -E^iE^j+\frac12\delta^{ij}\vec{E}{ }^2$ in the
electrostatic case,
see footnote~\ref{Footnote-sign-stress-tensor} on sign conventions,
explains the sign pattern of $s(r)$ and $p(r)$ at long distances,
where the electric field in the classical proton model corresponds 
to that of a point charge 
$E^i=\frac{e}{4\pi}\,\frac{r^i}{r^3}$ and implies asymptotically
$s(r) = -\,\frac{\alpha}{ 4\pi}\,\frac{\hbar c}{r^4}$ and
$p(r) =    \frac{\alpha}{24\pi}\,\frac{\hbar c}{r^4}$.

This has several consequences. First, the radial pressure $p_r(r)$ 
in the classical proton model does not comply with the condition 
\eqref{Eq:p-radial} for $r>2\,\rm fm$, but this is irrelevant 
as argued in~\cite{Varma:2020crx}. 
The point is that the mechanical system is confined within the radius
$R_{\rm dust}\simeq1\,\rm fm$, where the matter (dust) sits in the classical 
model. Here the mechanical stability criterion \eqref{Eq:p-radial}
should hold and does so. At distances $r>R_{\rm dust}$ we have only fields.
The change of sign of $p_r(r)$ in \eqref{Eq:p-radial} at $r\simeq2\,\rm fm$
{\it would} imply a mechanical instability and expel matter to infinity
{\it if} there was any at that point. But there is~none. Hence the system
is mechanically stable. In fact, the~stability of the proton in the 
classical model was proven in~\cite{Bialynicki-Birula:1993shm}.
Within a classical field theoretical framework, such an argumentation
is physically sound~\cite{Varma:2020crx}.\footnote{How one could transpose such a reasoning to a quantum field theoretical setting is an interesting question which remains to be elaborated.}

The second consequence is that the $D$-term of the proton is undefined. 
With $s(r)$ and $p(r)$ exhibiting $\frac1{r^4}$-tails at long-distances, 
the integrals in Eqs.~(\ref{Eq:D-from-s},~\ref{Eq:D-from-p}) diverge. 
The form factor $D(t)$, however, is well-defined for $t<0$ and given by
$D(t) = \frac{\alpha\pi}{4}\,\frac{m}{\sqrt{-t}}\,+\:$subleading terms
for $(-t)\ll m^2$~\cite{Varma:2020crx}.

The impact of QED effects on $D(t)$ was actually known in 
literature from chiral perturbation theory studies of charged pion EMT 
properties~\cite{Kubis:1999db} and other contexts~\cite{Donoghue:2001qc}, and 
further investigated subsequently in QED~\cite{Metz:2021lqv,Freese:2022jlu}.
It is not surprizing that a classical model can correctly reproduce
QED results~\cite{Varma:2020crx},
because classical electrodynamics is the long-distance limit of QED~\cite{Donoghue:2001qc}. 
Can this QED effect be seen experimentally? The answer is no. 

The QED asymptotics $D(t) = \frac{\alpha\pi}{4}\,\frac{m}{\sqrt{-t}}$, 
where the $D$-term form factor is positive, becomes apparent 
in the classical model for 
$(-t)\lesssim 10^{-5}\,\rm GeV^2$, which is 
beyond the reach of realistic experiments~\cite{Burkert:2018bqq}.
In the region $10^{-2}\,\rm GeV^2\lesssim(-t)\lesssim 1\,{\rm GeV}^2$, the 
proton $D(t)$ is negative and well described in the classical model by 
a multipole form $D(t)=D_{\rm reg}/(1-t/m_D^2)^n$, where $n\simeq(\mbox{2--3})$
and $D_{\rm reg}$ is a ``regularized value'' of the $D$-term which can be 
derived within the classical model in a ``regularization procedure'' that effectively removes the contribution from the classical $\frac{1}{r^4}$-tail~\cite{Varma:2020crx}. This is the behavior of the proton $D$-term one will be able
to infer experimentally.

The introduction of massless degrees of freedom like photons\footnote{The classical model can be used again for pedagogical purposes:
    the contributions of the massive carriers of the strong scalar and vector forces
    to, e.g., $s(r)$ are proportional to $\frac{(1+m_i r){ }^2}{r^4}\,e^{-2m_i r}$ for 
    $r>R_{\rm dust}$. In the limit $m_i\to0$ we recover the $\frac1{r^4}$-behavior of the massless Coulomb field, which makes the $D$-term in Eq.~\eqref{Eq:D-from-s} undefined.}
in a theory affects also other EMT form factors. But the values $A(0)=1$ and $J(0)=\frac12$, 
in the proton case, are fixed by Poincar\'e invariance and the QED effects manifest themselves
only in the slopes of these form factors~\cite{Donoghue:2001qc}.
Once again, the form factor $D(t)$ appears as the proton property 
most sensitive to changes in the dynamics. Julia Panteleeva has shown that these issues are not specific 
to QED and that $D$-terms are undefined also in other theories with 
long-range forces~\cite{Panteleeva:2023aiz}.

\newpage
\section{Thoughts on $D$-terms in atomic systems}
\label{Sec-11:D-in-atoms}

Atoms, and especially the exactly solvable hydrogen atom, 
provide an appealing theoretical framework for investigating the 
EMT interpretation.  
This has been pursued with the interesting finding that the $D$-term 
of the hydrogen atom is positive~\cite{Ji:2021mfb,Ji:2022exr,Czarnecki:2023yqd,Freese:2024rkr},
at variance with 
the findings of negative $D$-terms obtained in systems governed by short-range forces,
see Sec.~\ref{Sec-9:D-conjecture}.
Can one reconcile these different behaviors?
The cautious answer at this point is: possibly yes, though more work may be
needed to settle the issue.

First, it is important to stress that a different behavior of atomic $D$-terms 
may not be entirely surprising because atoms are bound states of the 
electromagnetic interaction. In fact, we learned in Sec.~\ref{Sec-10:em-effects} 
that the presence of long-range forces has drastic effects and generates,
e.g., due to long-distance behavior of the electric field $\propto\frac1{r^2}$
for the proton, a behavior like 
$D(t) \to \frac{\alpha\pi}{4}\,\frac{m}{\sqrt{-t}} \to +\,\infty$
as $|t|\to 0$, see Sec.~\ref{Sec-10:em-effects}.

This universal behavior is, of course, the same for proton and electron.
In the hydrogen atom, the proton charge is screened by the electron cloud and
the system is overall neutral. The QED singularities due to the proton and
electron charges compensate, and $D(t)$ is finite.
Although the hydrogen atom is exactly solvable, the calculation of its 
$D$-term is too technical to reproduce in a few lines, but the positive sign
can easily be explained. The ground-state wave function of the
electron in the hydrogen atom is $\Psi(r)=e^{-r/a}/\sqrt{\pi a^3}$, where 
$a$ is the Bohr radius. The ``charge density'' of the electron
cloud is $\rho_e(r)=-\,e\,|\Psi(r)|^2$. At distances $r$ much larger than
the proton radius $R_p$, the electric field of the proton-electron-cloud system is 
$E^i = \frac{e}{4\pi}(\frac1{r^2}+\frac{2}{ar}+\frac{2}{a^2})\,\frac{r^i}{r}\,e^{-2r/a}$. From $T^{ij}=-E^iE^j+\frac12\delta^{ij}\vec{E}{ }^2$, we then see
that $p(r)$ is positive and $s(r)$ is negative at asymptotic distances, i.e.\ the
same sign pattern emerges as in Sec.~\ref{Sec-10:em-effects}, except that now both 
functions are suppressed as $e^{-4r/a}$ at long distances. This makes the 
integrals (\ref{Eq:D-from-s},~\ref{Eq:D-from-p}) convergent,
and we obtain $0< D < \infty$. This back-of-the envelope estimate is in line with
the more involved calculations of Refs.~\cite{Ji:2021mfb,Ji:2022exr,Czarnecki:2023yqd,Freese:2024rkr}. 

However, we should keep in mind that the main point of Maxim's pressure interpretation
is not that the $D$-term is negative, but that the system can be treated as a
mechanical continuum system (and only if it can, then the mechanical stability 
criteria can be applied and imply that $D<0$~\cite{Perevalova:2016dln}). 
In order to keep the discussion simple, let us have in mind for the following
a comparison of the $D$-terms of a neutron and hydrogen atom. Both are
electrically neutral systems and have well-defined finite $D$-terms, 
but one has a negative and the other a positive $D$-term. 

One possible approach to put the positive $D$-terms of atoms into 
context is to rephrase our question as: can the interiors of atoms and hadrons
both be reasonably treated as mechanical continuum systems? 

Arguably, atoms and hadrons are very different systems and are not equally 
amenable to a mechanical continuum interpretation~\cite{Burkert:2023wzr}. Hadrons are extremely dense
systems. In the case of the proton, the density is of the order of $1\,\rm GeV/fm^3$.
The dynamics of this dense medium inside hadrons is governed by the extremely strong QCD forces. 
It is an assumption that mechanical continuum concepts can be applied to hadrons,
but an assumption which is physically not unreasonable.

We need to contrast this with the ``medium'' inside the hydrogen atom. The ``medium''
here is basically the electron cloud. Hence, we deal with an extremely dilute system with a density of one
electron mass distributed over a volume of $1\,\mbox{\AA}{ }^3$, i.e.\ the medium
in the interior of an atom is 18 orders of magnitude more dilute compared to the
medium inside a hadron. 

Let's also consider ``what kind of medium'' we deal with inside a hydrogen atom.
The hydrogen atom is a two-body system with a pointlike, infinitely positive charge
in the center which binds the electron. Thus the ``medium'' is actually one single
body, the electron, which swirls around. The quantum mechanical solution determines
the probability where the electron can be found. Does this constitute a physically
viable ``mechanical continuum system''? 
Notice, that even if we consider atoms in the large-$Z$ limit, then there still remain 
fundamental differences to, e.g., baryons in the large-$N_c$ limit~\cite{Diakonov:1996sr}.

To conclude this section, one possible explanation why we cannot compare neutral hadrons
and atoms is because we might be comparing apples and oranges. This is an admittedly
intuitive approach. More work may be needed to fully understand why, in
the context of the $D$-term, atoms may not provide useful toy model systems to get
insights about the dynamics inside hadrons.\footnote{For completeness, let us mention that positive $D$-terms have systematically been found in the covariant quark-diquark model calculations of Refs.~\cite{Fu:2022rkn,Fu:2023ijy,Wang:2023bjp,Wang:2024abv}. A positive $D$-term has also been found in the scalar diquark model calculation of Ref.~\cite{Amor-Quiroz:2023rke}, but only in the unrealistic case where the parent hadron and the active quark have the same mass. Depending on the modeling of the effective nucleon-quark-diquark vertex, diquark approaches do not always describe the nucleon as a dynamical quark-diquark bound state compliant with the virial theorem. Often additional model assumption are needed to obtain a negative $D(t)$~\cite{Muller:2014tqa}. Notice that the compliance of the pressure obtained from $D(t)$ with the von Laue condition does not demonstrate the stability of the system because {\it any} $D(t)$ satisfying $\lim_{t\to 0}tD(t)=0$ (and such that the integral in Eq.~\eqref{BFdef} exists) yields a pressure satisfying the von Laue condition~\cite{Hudson:2017xug}.}

\newpage

\section{Conclusions and outlook}
\label{Sec-5:conclusions}

Interpreting hadronic matrix elements of the stress tensor in terms of the concepts of ``pressure'' and ``shear forces'' borrowed from continuum mechanics is far from a trivial step. In this contribution, we reviewed the interpretation proposed by Maxim Polyakov in the seminal paper~\cite{Polyakov:2002yz} and addressed the criticism raised in the recent literature~\cite{Jaffe:2020ebz,Freese:2021czn,Ji:2021mtz,Ji:2021mfb,Ji:2022exr,Fujita:2022jus,Czarnecki:2023yqd}. 

We showed that the concept of pressure is not limited to the familiar situation found in homogeneous gases. Mechanical \textit{equilibrium} requires that the spatial integral of pressure must vanish for a bound system, which is basically the physical content of the virial theorem. This implies that pressure is not always positive and isotropic. A bound system results from the balance between repulsive and attractive forces.

Mechanical \textit{stability} requires, however, stronger constraints. Based on the analogy with continuum mechanics, it has been argued that a stable system should be characterized by a negative $D$-term~\cite{Perevalova:2016dln}, in agreement with results from hadronic models, lattice QCD, and dispersion relation studies~\cite{Burkert:2023wzr}. 

The fact that the $D$-term is found positive in the hydrogen atom has, however, led some authors to conclude that the above mechanical interpretation does not make sense~\cite{Ji:2021mtz,Ji:2021mfb,Ji:2022exr,Czarnecki:2023yqd}. We believe that this conclusion may be premature. We stressed, in particular, that a negative $D$-term is a natural consequence of mechanically stable systems determined by short-range forces, whereas the structure of atoms is governed by the long-range electromagnetic forces. 

While clarifying the domain of validity of the stability criterion requires more work, we conclude that none of the raised criticism to date has really invalidated Maxim's mechanical interpretation of the hadronic stress tensor.

\ \\
{\bf Acknowledgments.} 
This work was supported by the National Science Foundation under Awards 2111490 and 2412625, and the U.S. Department of Energy under the umbrella of the Quark-Gluon Tomography (QGT) Topical Collaboration with Award No. DE-SC0023646.

\end{document}